\documentstyle[]{article}
\def\bea{\begin{eqnarray}}
\def\eea{\end{eqnarray}}
\def\nn{\nonumber}

\def\beq{\begin{equation}}
\def\eeq{\end{equation}}
\def\ba{\beq\new\begin{array}{c}}
\def\ea{\end{array}\eeq}
\def\be{\ba}
\def\ee{\ea}

\def\Tr{{\rm Tr}}

\makeatletter
\newdimen\normalarrayskip              
\newdimen\minarrayskip                 
\normalarrayskip\baselineskip
\minarrayskip\jot
\newif\ifold             \oldtrue            \def\new{\oldfalse}
\def\arraymode{\ifold\relax\else\displaystyle\fi} 
\def\eqnumphantom{\phantom{(\theequation)}}     
\def\@arrayskip{\ifold\baselineskip\z@\lineskip\z@
     \else
     \baselineskip\minarrayskip\lineskip2\minarrayskip\fi}
\def\@arrayclassz{\ifcase \@lastchclass \@acolampacol \or
\@ampacol \or \or \or \@addamp \or
   \@acolampacol \or \@firstampfalse \@acol \fi
\edef\@preamble{\@preamble
  \ifcase \@chnum
     \hfil$\relax\arraymode\@sharp$\hfil
     \or $\relax\arraymode\@sharp$\hfil
     \or \hfil$\relax\arraymode\@sharp$\fi}}
\def\@array[#1]#2{\setbox\@arstrutbox=\hbox{\vrule
     height\arraystretch \ht\strutbox
     depth\arraystretch \dp\strutbox
     width\z@}\@mkpream{#2}\edef\@preamble{\halign
\noexpand\@halignto
\bgroup \tabskip\z@ \@arstrut \@preamble \tabskip\z@ \cr}%
\let\@startpbox\@@startpbox \let\@endpbox\@@endpbox
  \if #1t\vtop \else \if#1b\vbox \else \vcenter \fi\fi
  \bgroup \let\par\relax
  \let\@sharp##\let\protect\relax
  \@arrayskip\@preamble}
\def\eqnarray{\stepcounter{equation}%
              \let\@currentlabel=\theequation
              \global\@eqnswtrue
              \global\@eqcnt\z@
              \tabskip\@centering
              \let\\=\@eqncr
              $$%
 \halign to \displaywidth\bgroup
    \eqnumphantom\@eqnsel\hskip\@centering
    $\displaystyle \tabskip\z@ {##}$%
    \global\@eqcnt\@ne \hskip 2\arraycolsep
         $\displaystyle\arraymode{##}$\hfil
    \global\@eqcnt\tw@ \hskip 2\arraycolsep
         $\displaystyle\tabskip\z@{##}$\hfil
         \tabskip\@centering
    &{##}\tabskip\z@\cr}
\newfont{\hr}{msbm10}
\newfont{\ams}{msam10}

\begin{document}

\begin{titlepage}
\setcounter{footnote}0
\begin{center}
\hfill ITEP/TH-6/96\\
\hfill FIAN/TD-5/96\\
\hfill hepth/9603140\\
\vspace{0.3in}
{\LARGE\bf ${\cal N}=2$ Supersymmetric QCD and
       Integrable Spin Chains: Rational Case
N$_f<$ 2N$_c$}
\\
\bigskip
\bigskip
\bigskip
{\Large A.Gorsky
\footnote{E-mail address: gorsky@vxitep.itep.ru, sasha@rhea.teorfys.uu.se}
$^{\dag}$,
A.Marshakov
\footnote{E-mail address:
mars@lpi.ac.ru, andrei@rhea.teorfys.uu.se, marshakov@nbivax.nbi.dk}$^{\ddag}$,
A.Mironov
\footnote{E-mail address:
mironov@lpi.ac.ru, mironov@grotte.teorfys.uu.se}$^{\ddag}$,
A.Morozov
\footnote{E-mail address:
morozov@vxdesy.desy.de}
$^{\dag}$}
\\
\bigskip
$\phantom{gh}^{\dag}${\it ITEP, Moscow, 117 259, Russia}\\
$\phantom{gh}^{\ddag}${\it Theory Department,  P. N. Lebedev Physics
Institute , Leninsky prospect, 53, Moscow,~117924, Russia\\
and ITEP, Moscow 117259, Russia}
\end{center}
\bigskip \bigskip

\begin{abstract}
The form of the spectral curve for $4d$ ${\cal N}=2$
supersymmetric Yang-Mills
theory with matter fields in the fundamental representation
of the gauge group
suggests that its $1d$ integrable counterpart should be looked for
among (inhomogeneous) $sl(2)$ spin chains with the length of the chain
being equal to the
number of colours $N_c$. For $N_f < 2N_c$ the relevant spin chain
is the simplest $XXX$-model, and this identification
is in agreement with the known results in Seiberg-Witten theory.
\end{abstract}

\end{titlepage}

\newpage
\setcounter{footnote}0

\section{Introduction}

Exact form of the abelian low-energy effective actions and BPS massive
spectra for the $4d$
${\cal N}=2$ SUSY Yang-Mills (SYM) theories
\cite{SW1}-\cite{5} possess a concise
description in terms of $1d$ integrable systems \cite{GKMMM}. The
{\it reasons} for this identity remain somewhat obscure, but
{\it the fact} itself is already well established \cite{6}-\cite{AN}.
To be precise, the relation between the Seiberg-Witten solutions (SW) and
integrable theories
was so far described in full detail only for  particular family of
models: the ${\cal N}=2$ SYM theory with one ($N_a=1$)
``matter" ${\cal N}=2$ hypermultiplet in the adjoint representation of
the gauge group $G$ -- which is known to be related to the
Calogero-Moser family of integrable systems
\cite{7,IM3}.
When the hypermultiplet decouples (its mass becomes infinite), the dimensional
transmutation takes place and the pure gauge $4d$ ${\cal N}=2$ SYM theory gets
associated with the Toda-chain model.
A physically more interesting family of models --
the ${\cal N}=2$ supersymmetric QCD (SQCD) with $N_f$ matter
${\cal N}=2$ hypermultiplets in {\it the fundamental} representation of
$G$ -- does not have a well established integrable counterpart yet. It is
known only \cite{Mar},\cite{AN} that the
$N_c = 3$, $N_f = 2$ curve can be associated with the
Goryachev-Chaplygin top.
The purpose of this letter is to fill, at least partly, this gap.  Our
suggestion is to associate the family of ${\cal N}=2$ SQCD models with the
well-known family of integrable systems -- (inhomogeneous $sl(2)$) spin
chains, of which the Toda chain (pure gauge model) is again a limiting case.
The crucial motivation for {\it such} a suggestion \cite{Mar}
is the peculiar form of the spectral equations, derived in \cite{SW2},
\cite{4}.
In this letter we just describe the idea, illustrating it by the simplest
example of the rational $XXX$ spin chain, which is, however,
enough for the complete
description of the $N_f < 2N_c$ case. The detailed arguments and
analysis of the most interesting elliptic case of $N_f = 2N_c$ are
postponed to a separate paper.

The SW problem is described by the following set of data
(see \cite{IM3} for details). Let us assume that the YM theory
is {\it softly} regularized both in the UV and IR regions -- this
is always possible in the ${\cal N}=2$ SUSY framework.
In the UV region,
the theory is embedded -- by addition of appropriate massive
matter ${\cal N}=2$ hypermultiplets -- into an UV-{\it finite} model.
At ultra-high energies, this {\it non-abelian} theory has
vanishing $\beta$-function, i.e. is conformally-invariant,
and possesses
a single coupling constant $\tau = \frac{4\pi i}{e^2} +
\frac{\theta}{2\pi}$. At the energies below the masses $m_\alpha$
of the additional matter hypermultiplets, the original ${\cal N}=2$ SUSY theory
is reproduced, which is thus labeled by the set of data
$\{G,\tau,m_\alpha\}$.

In the IR region, the theory can avoid entering the strong-coupling regime,
if the scalar components of the gauge supermultiplet develop non-zero vacuum
expectation values along the valleys of the superpotential.
These v.e.v.'s $\langle \Phi \rangle$ are given by diagonal
matrices and can be fully described by the set of ``moduli"
$h_k = \frac{1}{k}\langle \Tr\Phi^k \rangle$.
At energies below this IR ``soft cutoff", the theory becomes
${\cal N}=2$ SUSY {\it abelian} model, with the {\it set} of coupling
constants $T_{ij}$. $T_{ij}$ is actually expressed in terms of
``periods" $a^i$, $a_i^D=\frac{\partial{\cal F}}{\partial a^i}$:
$T_{ij} = \frac{\partial^2{\cal F}}{\partial a^i\partial a^j} =
\frac{\partial a^D_i}{\partial a^j}$.

The SW problem can be formally defined as a map
\be
G,\tau,m_\alpha,h_i \rightarrow T_{ij}, \ a^i,\ a_i^D
\ee
and the solution to this problem has an elegant description in the following
terms \cite{SW1,SW2}:
one associates with the data $G,\tau,m_\alpha$ a family of $2d$
surfaces (complex curves) ${\cal C}$ with $h_i$ parameterizing (some) moduli
of their complex structures, and a meromorphic 1-form
$dS$ on every ${\cal C}$. Then
$a^i = \oint_{A_i} dS$, $a_i^D = \oint_{B^i} dS$.
In terms of integrability theory the curves ${\cal C}$ are
interpreted \cite{GKMMM} as the spectral curves of certain
integrable systems, and $a^i$, $a_i^D$ are related to the action integrals
($\oint pdq$) of the system. Thus, to describe the solution to the SW problem
one should present the explicit map
\be
G, \tau, m_\alpha \rightarrow \left( {\cal C},\ dS
\right)\{h_i\},
\ee
and this turns out to be equivalent to selection of particular integrable
system. The bare charge $\tau $ disappears from the formulas in the
asymptotically free region $N_f < 2N_c$, where dynamical
transmutation substitutes $\tau$ by $\Lambda _{QCD}^{(N_f)} \sim
\exp \frac{2\pi i\tau}{2N_c-N_f}$. In what follows
we put $\Lambda _{QCD}^{(N_f)} = 1$.

\section{From Toda to Spin Chains}

Our starting point is that the Toda chain spectral curves, corresponding
to the pure gauge ($N_f=0$) ${\cal N}=2$ SUSY theory \cite{GKMMM},
can be described in terms
of two {\it different} characteristic equations. The first one,
\be\label{fscToda}
\det_{N_c\times N_c}
\left({\cal L}^{\rm TC}(w) - \lambda\right) = 0,
\ee
with $N_c\times N_c$ matrix ${\cal L}^{\rm TC}(w)$
being the Lax operator of the
periodic Toda chain, can be obtained from a degeneration of the elliptic
Calogero-Moser particle system, and this fact is crucially used
in description of models with adjoint matter hypermultiplet.

Equation (\ref{fscToda}) reads
\be\label{fsc-Toda}
w + \frac{1}{w} = 2P_{N_c}(\lambda),
\ee
due to the very particular form of the matrix
${\cal L}^{\rm TC}(w)$ (not
preserved by its Calogero-Moser generalization).
Here $P_{N_c}(\lambda )$ is a polynomial of degree $N_c$,
whose coefficients are the Schur polynomials of the Toda chain Hamiltonians
$h_k = \sum_{i=1}^{N_c} p_i^k + \ldots$:
\be
P_{N_c}(\lambda ) =  \sum_{k=0}^{N_c}
{\cal S}_{N_c-k}(h) \lambda ^{N_c} = \nn \\
= \left( \lambda ^{N_c} + h_1 \lambda ^{N_c-1} +
\frac{1}{2}(h_2-h_1^2)\lambda ^{N_c-2} + \ldots \right).
\ee
Since (\ref{fsc-Toda}) is quadratic equation with
respect to $w$, one can rewrite it as another characteristic equation
involving only $2\times 2$ matrices
\be\label{mon}
\det _{2\times 2}\left( T_{N_c}(\lambda ) - w \right) =
w^2 - w\ \Tr\ T_{N_c}(\lambda ) + \det T_{N_c}(\lambda ) = 0.
\ee
In the Toda-chain case, the $2\times 2$ matrix $T_{N_c}(\lambda)$ is such that
$\Tr\ T_{N_c}^{\rm TC}(\lambda ) = P_{N_c}(\lambda )$ and
$\det T_{N_c}^{\rm TC}(\lambda ) = 1$.
According to \cite{SW2}, \cite{4}, the spectral curves for
the ${\cal N}=2$ SQCD with any $N_f < 2N_c$ have the same
form (\ref{mon}) with
\be\label{trdet}
\Tr\ T_{N_c}(\lambda) = P_{N_c}(\lambda) + R_{N_c-1}(\lambda),
\ \ \ \ \det T_{N_c}(\lambda ) = Q_{N_f}(\lambda ),
\ee
and $Q_{N_f}(\lambda)$ and $R_{N_c-1}(\lambda)$ are certain
$h$-{\it independent} polinomials of $\lambda$.

To go further, let us remind the origin of representation (\ref{mon})
for the Toda-chain theory. The $N_c\times N_c$ Lax equation
${\cal L}_{ij}\psi _j = \lambda\psi _i$ can be rewritten
through $2\times 2$ matrices \cite{FT}:
\be\label{LaxToda}
{\tilde\psi}_{i+1} = L_i^{\rm TC} (\lambda ){\tilde\psi}_{i},
\nn \\
{\tilde\psi}_i =
\left(\begin{array}{c}
 \psi _{i} \\ \chi _i \end{array}\right) , \ \ \ \
L_i^{\rm TC}(\lambda) =
\left(\begin{array}{cc}
 p_i + \lambda & e^{q_i} \\
-e^{-q_i} & 0 \end{array}\right),
\ee
i.e. $\chi_{i+1}=-e^{-q_i}\psi_i$.
Eq.(\ref{mon}) is expressed through the monodromy matrix,
\be\label{Tmat}
T_{N_c}^{\rm TC}(\lambda ) = \prod _{i = N_c}^1 L_i^{\rm TC}(\lambda ), \ \ \
hbox{thus}\ \ \ T_{N_c}(\lambda ){\tilde\psi}_i = {\tilde\psi}_{i + N_c}
\ee
with $\det _{2\times 2} T_{N_c}^{\rm TC}(\lambda) =
\prod_{i=1}^{N_c} {\rm det}_{2\times 2} L_i^{\rm TC}(\lambda-\lambda_i) = 1$
and
$\Tr\  T_{N_c}^{\rm TC}(\lambda ) = P_{N_c}(\lambda )$.
Eq.(\ref{mon}) can be understood as a corollary of the boundary condition
$\tilde \psi_{i+N_c} = w\tilde\psi_i$. Substitution of (\ref{Tmat}) into
(\ref{mon}) gives rise to the Toda-chain spectral curve (\ref{fsc-Toda}).
Together with the formula for the 1-form $dS= \lambda\frac{dw}{w}$
this provides the solution to the SW problem for pure gauge ($N_f=0$) theory.

Thus, we reproduce the spectral curve (\ref{fsc-Toda})
and the 1-form $dS$ of the periodic
Toda-chain system from the different perspective -- taking a closed chain
(of length = $N_c$) of $2\times 2$ Lax matrices and computing the eigenvalues
of the monodromy operator.
The two descriptions, (\ref{fscToda}) and (\ref{mon}),
are identically equivalent for the Toda
chain, but their {\it deformations} are very different:
the ``chain" representation (\ref{mon}), (\ref{Tmat})
is naturally embedded into the family of
$XYZ$ spin chains \cite{FT,Skl},
while the $N_c\times N_c$ Lax operator representation --
into that of Calogero-Moser models and generic Hitchin systems
\cite{CalfroHit}. Our suggestion is to
associate these two different deformations of the integrable Toda chain
system with the two different deformations of the pure ${\cal N}=2$ SYM
theory: by addition of massive matter multiplets in the {\it fundamental} and
{\it adjoint} representations of the gauge group $G = SU(N_c)$ respectively.
Self-\-consistency of the $4d$ theory in the UV region requires that $N_f\leq
2N_c$ and $N_a\leq 1$, thus the numbers of deformation parameters (masses
$m_\alpha$) in the two cases are $2N_c$ and $1$. Since adjoint model
is exhaustively analyzed in \cite{7,IM3}, in what follows we concentrate
on the fundamental case.

Integrability of the Toda chain in
representation (\ref{LaxToda}) follows from
{\it quadratic} r-matrix relations \cite{Skl}
\be\label{quadr-r}
\left\{L(\lambda)\stackrel{\otimes}{,}L(\lambda')\right\} =
\left[ r(\lambda-\lambda'),\ L(\lambda)\otimes L(\lambda')\right],
\ee
so that $\{p_i,q_j\} = \delta_{ij}$ follows from
(\ref{quadr-r}) with the rational $r$-matrix (see (\ref{rat-r})
below).
The crucial property of this relation is that it
is multiplicative and any product like (\ref{Tmat})
satisfies the same relation
\be\label{Tbr}
\left\{T_{N_c}(\lambda)\stackrel{\otimes}{,}T_{N_c}(\lambda')\right\} =
\left[ r(\lambda-\lambda'),\
T_{N_c}(\lambda)\otimes T_{N_c}(\lambda')\right],
\ee
provided all $L_i$ in product (\ref{Tmat})
are independent, $\{L_i,L_j\} = 0$ for $i\neq j$.

Our proposal is to look at non-Hitchin generalizations of the Toda chain,
i.e. deform eqs.(\ref{mon})-(\ref{Tmat}) preserving the quadraticity of
Poisson brackets (\ref{Tbr}) and, thus, the possibility to build a
monodromy matrix $T(\lambda)$ by multiplication of $L_i(\lambda)$'s.
For a moment, we even allow $L(\lambda)$ to be $n\times n$, not
obligatory $2\times 2$ matrices.

The full spectral curve for the periodic {\it inhomogeneous} spin chain is
given by:
\be\label{fsc-SCh}
\det_{n\times n}\left(T_{N_c}(\lambda) -  w\right) = 0,
\ee
with the inhomogeneous $T$-matrix
\be\label{T-matrix}
T_{N_c}(\lambda) = \prod_{i=N_c}^1 L_i(\lambda-\lambda_i)
\ee
still satisfying (\ref{Tbr}), and $d^{-1}({\rm symplectic\ form})$ is now
\be\label{1f}
dS = \lambda\frac{d\tilde w}{\tilde w}, \nn \\
{\tilde w} = w\cdot (\det T_{N_c})^{-1/n}.
\ee
In the particular case of ${\cal N}=2$ ($sl(2)$ spin chains), the spectral
equation acquires the form (\ref{mon}) (in general the spectral
equation is of the $n$-th order in $w$):
\be\label{fsc-sc1}
w + \frac{\det_{2\times 2} T_{N_c}(\lambda)}{w} =
 \Tr_{2\times 2}T_{N_c}(\lambda ),
\ee
or
\be
{\tilde w} + \frac{1}{\tilde w} = \frac{\Tr_{2\times 2}T_{N_c}(\lambda)}
{\sqrt{\det_{2\times 2} T_{N_c}(\lambda)}}.
\label{fsc-sc2}
\ee
The r.h.s. of this equation contains the dynamical variables of
the spin system only in the special combinations -- its
Hamiltonians (which are all in involution, i.e. Poisson-commuting).
It is this peculiar shape (quadratic $w$-dependence) that suggests the
identification of the periodic $sl(2)$ spin chains with solutions to the
SW problem with the fundamental matter supermultiplets.

\section{$XXX$ Spin Chain and the Low Energy SYM with $N_f < 2N_c$}

The $2\times 2$ Lax matrix for the $sl(2)$ $XXX$ chain is
\be
L(\lambda) = \lambda \cdot {\bf 1} + \sum_{a=1}^3 S_a\cdot\sigma^a.
\ee
The Poisson brackets of the dynamical variables $S_a$, $a=1,2,3$
(taking values in the algebra of functions)
are implied by (\ref{quadr-r}) with the rational $r$-matrix
\be\label{rat-r}
r(\lambda) = \frac{1}{\lambda}\sum_{a=1}^3 \sigma^a\otimes \sigma^a.
\ee
In the $sl(2)$ case, they are just
\be\label{Scomrel}
\{S_a,S_b\} = i\epsilon_{abc} S_c,
\ee
i.e. $\{S_a\}$ plays the role of angular momentum (``classical spin'')
giving the name ``spin-chains'' to the whole class of systems.
Algebra (\ref{Scomrel}) has an obvious Casimir operator
(an invariant, which Poisson commutes with all the generators $S_a$),
\be\label{Cas}
K^2 = {\bf S}^2 = \sum_{a=1}^3 S_aS_a,
\ee
so that
\be\label{detTxxx}
\det_{2\times 2} L(\lambda) = \lambda^2 - K^2,
\nn \\
\det_{2\times 2} T_{N_c}(\lambda) = \prod_{i=N_c}^1
\det_{2\times 2} L_i(\lambda-\lambda_i) =
\prod_{i=N_c}^1 \left((\lambda - \lambda _i)^2 - K_i^2\right) = \nn \\
= \prod_{i=N_c}^1(\lambda + m_i^+)(\lambda + m_i^-)
= Q_{2N_c}(\lambda),
\ee
where we assumed that the values of spin $K$ can be different at
different nodes of the chain, and
\footnote{
Eq.(\ref{mpm}) implies that the limit of vanishing masses, all
$m_i^\pm = 0$, is associated with the {\it homogeneous} chain
(all $\lambda_i = 0$) and vanishing spins at each site (all $K_i = 0$).
It deserves noting that a similar situation was considered by
L.Lipatov \cite{L} in the study of {\it the high}-energy limit of
the ordinary (non-supersymmetric) QCD.
The spectral equation is then the classical limit of the
Baxter equation from \cite{FK}.  }
\be
m_i^{\pm} = -\lambda_i \mp K_i.
\label{mpm}
\ee
While the determinant of monodromy matrix (\ref{detTxxx})
depends on dynamical variables
only through Casimirs $K_i$ of the Poisson algebra, the dependence of
the trace
${\cal T}_{N_c}(\lambda) =
\frac{1}{2}\Tr_{2\times 2}T_{N_c}(\lambda)$ is less trivial.
Still, as usual for integrable systems, it depends
on $S_a^{(i)}$ only through Hamiltonians of the spin chain (which are not
Casimirs but Poisson-commute with {\it each other}).

In order to get some impression how the Hamiltonians
look like, we present explicit examples of monodromy matrices
for $N_c = 2$ and $3$. Hamiltonians depend non-\-trivially
on the $\lambda_i$-parameters (in\-homo\-geneities of the chain)
and the coefficients
in the spectral equation (\ref{fsc-SCh}) depend only on the
Hamiltonians and symmetric functions of the $m$-parameters (\ref{mpm}),
i.e. the dependence of $\{\lambda_i\}$ and $\{K_i\}$ is rather special.
This property is crucial for identification of the $m$-parameters
with the masses of the matter supermultiplets in the ${\cal N}=2$ SQCD.

\bigskip

\bigskip

\noindent
{\bf N$_c=$ 2}\\
\be
{\cal T}_2(\lambda) =  (\lambda - \lambda_1)(\lambda - \lambda_2) -
\sum_{a=1}^3 S_a^{(1)}S_a^{(2)} = \nn \\ =
\lambda^2 - (\lambda_1 + \lambda_2)\lambda +
\left(h_2 + \lambda_1\lambda_2 - \frac{1}{2}(K_1^2+K_2^2)
- \frac{1}{2}(\lambda_1^2 + \lambda_2^2)\right).
\label{calT2}
\ee
The second Hamiltonian is
\be
h_2\{\lambda_i\} = -\sum_{a=1}^3\sum_{i<j}^{N_c} S_a^{(i)}S_a^{(j)} -
\frac{1}{4}t_1(K^2) - \frac{1}{4}t_1(\lambda^2) = \nn \\
\stackrel{N_c=2}{=}
-\sum_{a=1}^3 S_a^{(1)}S_a^{(2)} - \frac{1}{4}(K_1^2+K_2^2) -
\frac{1}{4}(\lambda_1^2
+ \lambda_2^2).
\label{h2def}
\ee
The coefficient of the $\lambda^1$-term at the r.h.s. of (\ref{calT2})
can be expressed through the parameters $m^{\pm}$, defined by (\ref{mpm}):
\be
-(\lambda_1+\lambda_2) = \frac{1}{2}(m_1^+ + m_1^- + m_2^+ + m_2^-)
= \frac{1}{2}\sum_{\gamma =1}^{2N_c} m_{\gamma} = \frac{1}{2}t_1\{m\}.
\ee
where we introduced an obvious notation $\{m_\gamma\}$ for the whole
set of parameters $\{m^\pm_i\}$, and the symmetric functions
are defined as
\be
t_k\{m\} = \sum_{\gamma_1 < \ldots < \gamma_k} m_{\gamma_1}\ldots
m_{\gamma_k}
\ee
for any sets of variables.

The last ($\lambda^0$) term at the r.h.s. of (\ref{calT2})
can be represented as
\be
h_2 + \lambda_1\lambda_2 + \frac{1}{4}(K_1^2+K_2^2)
+ \frac{1}{4}(\lambda_1^2 + \lambda_2^2) = \nn \\
= h_2 + t_2(\lambda) + \frac{1}{4}t_1(K^2) + \frac{1}{4}t_1(\lambda^2) =
h_2\{\lambda_i\} + \frac{1}{4}t_2\{m\}.
\ee
Indeed,
\be
t_2\{m\} = \frac{1}{2}\left(\left(\sum_{\gamma =1}^{2N_c}
m_\gamma\right)^2 - \sum_{\gamma=1}^{2N_c} m_\gamma^2\right) = \nn \\ =
\frac{1}{2}\left( \left(2\sum_{i=1}^{N_c} \lambda_i\right)^2 -
2\sum_{i=1}^{N_c}\left(\lambda_i^2 - K_i^2\right)\right) =
4 t_2(\lambda) + t_1(\lambda^2) + t_1(K^2).
\ee

\bigskip

\bigskip

\noindent
{\bf N$_c=$ 3}\\
In this case:
\be
{\cal T}_3(\lambda) = \lambda^3 - (\lambda_1+\lambda_2+\lambda_3)\lambda^2
+ \nn \\ +
\left(\lambda_1\lambda_2 + \lambda_2\lambda_3 + \lambda_3\lambda_1
- \sum_{a=1}^3(S_a^{(1)}S_a^{(2)} + S_a^{(2)}S_a^{(3)} +
S_a^{(3)}S_a^{(1)})\right)\lambda\ + \nn \\ +
i\epsilon_{abc}S_a^{(1)}S_b^{(2)}S_c^{(3)} = \nn \\
= \lambda^3 + \frac{1}{2}t_1\{m\}\lambda^2 +
(h_2 + \frac{1}{4}t_2\{m\})\lambda + (h_3 + \frac{1}{8}t_3\{m\}),
\ee
where $h_2\{\lambda_i\}$ has been already defined in (\ref{h2def}) and
\be
h_3\{\lambda_i\} = i\epsilon_{abc}\sum_{i<j<k}^{N_c}
S_a^{(i)}S_b^{(j)}S_c^{(k)} +
\sum_i\sum_{\stackrel{j,k\neq i}{j<k}}\lambda_i
S_a^{(j)}S_a^{(k)} + \nn \\ +
\frac{1}{4}\left(
\left[t_1(\lambda^2) + t_1(K^2)\right] t_1(\lambda) -
t_1(\lambda^3) - t_1(\lambda K^2)\right),
\ee
while
\be
t_3\{m\} = \frac{1}{6}\left( \left(\sum m\right)^3 -
3\left(\sum m^2\right)\left(\sum m\right) + 2\sum m^3\right) = \nn \\ =
-8t_3(\lambda) - 2t_1(\lambda^2)t_1(\lambda) - 2t_1(K^2)t_1(\lambda)
+ 2t_1(\lambda^3) + 2t_1(\lambda K^2).
\ee

Similarly one can deduce that
\be
{\cal T}_{N_c}(\lambda) =
\frac{1}{2}\Tr_{2\times 2}T_{N_c}(\lambda) = P_{N_c}(\lambda |h) +
\sum \lambda ^{N_f - N_c - i}t_i\{ m\} = P_{N_c}(\lambda|h) +
R_{N_c-1}(\lambda|m).
\ee
Together with (\ref{1f})-(\ref{fsc-sc2}) and
(\ref{detTxxx}), this reproduces the
formulas proposed in \cite{4}.

Thus, we demonstrated that the SW problem for the
${\cal N}=2$ SUSY QCD with
$N_f < 2N_c$ is solved in terms of integrable $XXX$ spin chain.
This construction has a natural elliptic generalization, which
describes the conformal point
$N_f = 2N_c$. The details will be presented elsewhere.

\section{Acknowledgements}

We are indebted to
O.Aharony, A.Hanany, H.Itoyama, S.Kharchev, G.Korchemsky,
I.Krichever,N.Nekrasov I.Polyubin, V.Rubtsov, J.Sonnenschein and A.Zabrodin
for usefull discussions.

The work of
A.G. was partially supported by the grants RFFI 96--01-01101 and
INTAS 1010--CT93--0023;
that of A.Mar. -- by the RFFI 96--01-00887 and INTAS 93--06; and that of
A.Mir. by RFFI 96--01-01106 and INTAS 93--1038.
A.Mir. also acknowledges the support of the Volkswagen Stiftung project
``Integrable models and strings".

\end{document}